\documentclass[aps,prl,twocolumn]{revtex4}
\usepackage{array}
\usepackage{amsmath}
\usepackage{graphicx}
\usepackage{amssymb}
\usepackage{latexsym}

\newcommand{\bra}[1]{\ensuremath{\langle {#1} |}}
\newcommand{\ket}[1]{\ensuremath{| {#1} \rangle}}
\newcommand{\braket}[2]{
	\ensuremath{\langle {#1} | {#2} \rangle}}

\renewcommand{\vec}[1]{\mathbf{#1}}
\newcommand{\mat}[1]{\mathsf{#1}}
\newcommand{\op}[1]{\ensuremath{\hat{\mathbf{#1}}}}

\newcommand{\mysec}[1]{\emph{#1} ---}
\newcommand{\eqnref}[1]{Eq.~(\ref{#1})}

\begin{document}

\title{Analytic Time Evolution, Random Phase Approximation, and Green Functions
\\ for Matrix Product States}

\author{Jesse M. Kinder}
\author{Claire C. Ralph}
\author{Garnet Kin-Lic Chan}
\email{gc238@cornell.edu}
\affiliation{Department of Chemistry and Chemical Biology, Cornell University,
Ithaca, New York 14853}

\begin{abstract}

	Drawing on similarities in Hartree-Fock theory and the theory of matrix
	product states (MPS), we explore extensions to time evolution, response
	theory, and Green functions. We derive analytic equations of motion for MPS
	from the least action principle, which describe optimal evolution in the
	small time-step limit. We further show how linearized equations of motion
	yield a MPS random phase approximation, from which one obtains response
	functions and excitations. Finally we analyze the structure of site-based
	Green functions associated with MPS, as well as the structure of
	correlations introduced via the fluctuation-dissipation theorem.

\end{abstract}

\date{\today}

\maketitle

Matrix product states are a powerful class of quantum states for one dimensional
correlated quantum problems. They underpin the density matrix renormalization
group algorithm which has yielded unprecedented accuracy in many systems
\cite{white1993dmrg,rommer1997mps}. In an MPS wave function, the quantum state is
approximated by a contraction of independent tensors, each associated with a
site on the lattice. Because of the contracted product structure, it can be
viewed as a site-based mean field theory with finite entanglement.

A different class of mean field theories is based on independent particles. For
fermions, this is Hartree-Fock (HF) theory, where the quantum state is
approximated by a Slater determinant of independent electron orbitals. HF
theory is the foundation on which many other approximations are built. For
example, time-dependent Hartree-Fock theory (TDHF) approximates the true
evolution of a quantum state with a single Slater determinant. Linearizing the
equations of motion yields the random phase approximation (RPA) which gives the
excitation spectrum and response functions \cite{mclachlan1964tdhft}. Moreover,
the independent particle picture contained in HF theory is the starting point
for the Green function approach to many particle quantum systems.

Recently, there has been much work to generalize MPS beyond the ground state
formalism to time dependence \cite{xiang2003tddmrg, vidal2004tebd,
white2004tdmrg, daley2004tdmrg, schmitteckert2004net, verstraete2004tmps,
schollwock2010dmrg} and response properties \cite{hallberg1995dma,
kuhner1999dcf, jeckelmann2002ddmrg, dorando2009art}. Despite the different
physical content of the site-based mean field of the MPS and independent
particle mean-field of HF, similar mathematical structures imply that extensions
to MPS can be developed in analogy with HF theory. Here, we explore this
direction. First, we briefly summarize the HF and MPS approaches to ground state
energies, establishing our notation and the parallel structure of the theories.
We then derive analytic MPS equations of motion using the Dirac-Frenkel
variational principle. We show that they provide an optimal representation of
MPS time evolution, and discuss their relationship to time evolution schemes
currently in use. Next, we linearize the time evolution to derive an MPS analog
of the RPA, and show how excitation energies and response properties are
obtained through a linear eigenvalue problem. Finally, we explore the structure
of Green functions that emerge in the MPS theory, and the correlations that
arise in MPS through the fluctuation-dissipation theorem.

\mysec{Stationary States} In HF theory, the $N$-particle wave function is
approximated by a single Slater determinant of orbitals $\phi_i(\vec{r})$. The
variational objects of the theory are the orbitals, and the best mean-field
approximation to the ground state is the set of orbitals that minimizes the
expectation value of the Hamiltonian, $E$. This leads to the Hartree-Fock
equations, $\op{f} \, \phi_i(\vec{r}) = \varepsilon_i \phi_i(\vec{r})$, where
$\op{f}$ is the Fock operator. Thus, each orbital is an eigenstate of a
one-particle Hamiltonian that depends on all of the orbitals. In this sense, HF
theory is a mean field theory for independent particles. If the orbitals are
expanded in a finite basis, then the Hartree-Fock equations may be expressed in
matrix form:
\begin{equation}
	\mat{F} \cdot \mat{C} = E \, \mat{S} \cdot \mat{C} ,
	\label{eq:fock_matrix}
\end{equation}
where $\mat{F}$ is the Fock matrix, $\mat{S}$ is the overlap matrix, and
$\mat{C}$ is the coefficient matrix of the orbitals in the finite basis. Since
$\mat{F}$ is a function of $\mat{C}$, this equation must be solved
self-consistently.

The defining equations of an optimal MPS wave function can be derived in an
analogous way. In an MPS wave function for $k$ sites, the amplitude of a Fock
state is given by the trace of a product of matrices:
\begin{equation}
	\ket{\psi} = \sum_{\vec{n}} \text{Tr}
	\left( \prod_{i=1}^{k} \mat{A}_{i}^{n_i} \right)
	\ket{\vec{n}},
	\label{eq:mps}
\end{equation}
where $\ket{\vec{n}} = \ket{n_1,n_2,\dots}$ labels a state in Fock space. A
matrix $\mat{A}_{i}^{n_i}$ is associated with each site $i$ and orbital
occupancy, $n_i$. The size of the matrices defines the auxiliary dimension $M$,
and the maximum occupation number of a site defines the physical dimension $d$.
The $d$ independent $M \times M$ matrices associated with each site can be
collected into a single tensor, $(\mat{A}_i)^{n_i}_{\alpha \beta}$, where
$\alpha$ and $\beta$ are the auxiliary indices. This tensor can be flattened
into a vector labeled by a single compound index $I = (n_i,\alpha,\beta)$.
Derivatives with respect to the components of the tensor define a
(nonorthogonal) basis for the wave function:
\begin{equation}
	\ket{\psi_{I}}
	= \dfrac{ \partial \ket{\psi} }{ \partial \mat{A}_{i}^I }.
	\label{eq:basis}
\end{equation}
From the linear dependence of $\ket{\psi}$ on each matrix, we have $\ket{\psi} =
\sum_i \mat{A}_i^I \ket{\psi_{I}}$.

Requiring the variational energy to be stationary with respect to variations in
a single tensor defines an eigenvalue problem:
\begin{equation}
	\mat{H}_i \cdot \mat{A}_i = E \, \mat{S}_i \cdot \mat{A}_i .
	\label{eq:vmps}
\end{equation}
The matrices $\mat{H}_i$ and $\mat{S}_i$ are the Hamiltonian and overlap
matrices in the local basis defined by $\mat{A}_i$. Since the basis defined in
\eqnref{eq:basis} depends on the other tensors, $\mat{A}_i$ is the eigenvector
of an effective Hamiltonian defined by the other tensors, analogous to
\eqnref{eq:fock_matrix} for the HF orbitals.

The entire set of equations, one for each tensor, can be combined into a single
eigenvalue problem by defining the compound index $\mu = (i,n_i,\alpha,\beta)$
and collecting all the elements of all the tensors into a single vector
$\vec{A}$. This vector contains $k d M^2$ elements. The wave function can be
expanded as
\begin{equation}
	\ket{\psi} = \dfrac{1}{k} \sum_{\mu} A_\mu \ket{\psi_\mu} ,
\end{equation}
and the optimal MPS wave function is a self-consistent solution of
\begin{equation}
	\mat{H} \cdot \vec{A} = E \, \mat{S} \cdot \vec{A}.
\label{eq:mps_gs}
\end{equation}
The Hamiltonian and overlap matrices are defined as
\begin{align}
	\mat{H}_{\mu\nu}
	&= \dfrac{1}{k} \bra{ \psi_\mu } \mathcal{H} \ket{ \psi_\nu }
	\label{eq:eff_ham} \\
	\mat{S}_{\mu\nu}
	&= \dfrac{1}{k} \braket{ \psi_\mu }{ \psi_\nu } .
	\label{eq:overlap}
\end{align}

The power of MPS wave functions stems from their numerical efficiency. For MPS
with open boundary conditions (as used in the DMRG algorithm) expectation values
of local operators and Hamiltonians can be calculated with $\mathcal{O}(k M^3)$
complexity. The action of local operators and Hamiltonians, such as $\mat{H}
\cdot \vec{A}$ and $\mat{S} \cdot \vec{A}$, is also obtained with
$\mathcal{O}(k M^3)$ complexity. Solving the eigenvalue problem
\eqnref{eq:mps_gs} iteratively requires a cost proportional to the operations
$\mat{H} \cdot \vec{A}$ and $\mat{S} \cdot \vec{A}$ and is thus also of
$\mathcal{O}(k M^3)$ complexity. The prefactor depends on preconditioning for
both $\mat{H}$ and $\mat{S}$. The DMRG preconditions $\mat{S}$ by solving
\eqnref{eq:vmps} for a single tensor $\mat{A}_i$ at a time, where the overlap
$\mat{S}_i$ can be exactly removed by canonicalization.

\mysec{Time Evolution} The time-dependent Schr\"{o}dinger equation can be
derived by minimizing the Dirac-Frenkel action
\begin{equation}
	S =	\int dt \Big(
		i \hbar \, \bra{\psi} \dot{\psi} \rangle
		- \bra{\psi} \mathcal{H} \ket{\psi}
	\Big)
\end{equation}
with respect to arbitrary variations $\bra{\delta \psi}$. When the wave
function is constrained to a particular form, such as a Slater determinant or an
MPS, variations may only be taken with respect to the parameters $\{ \lambda_i
\}$ of the wave function, 
\begin{equation}
	\bra{\delta\psi} = \sum_i \delta \lambda_i \cdot \bra{ \partial_i \psi }
\end{equation}
The time derivative of $\ket{\psi}$ is also constrained in this way. Minimizing
the action then gives the best approximation to the true evolution of the wave
function \emph{within the space of variational wave functions}.

When applied to Hartree-Fock theory, the resulting equations of motion are the
time-dependent Hartree-Fock equations. The time-dependent version of
\eqnref{eq:fock_matrix} is
\begin{equation}
	 i \hbar \, \mat{S} \cdot \dfrac{d \mat{C}(t)}{dt}
	 = \mat{F}(t) \cdot \mat{C}(t).
	\label{eq:tdhft}
\end{equation}
The Fock matrix depends on $t$ through the time-dependence of the orbitals. It
may also depend explicitly on $t$ through a time-dependent external potential

Applying the same variational approach to the action of an MPS wave function
gives an equation of motion for the time dependence of the matrix elements:
\begin{equation}
	i \hbar \, \mat{S}(t) \cdot \dfrac{d \vec{A}(t)}{dt}
	= \mat{H}(t) \cdot \vec{A}(t) ,
	\label{eq:mps_dynamics}
\end{equation}
where $\mat{H}$ and $\mat{S}$ are the time-dependent versions of
\eqnref{eq:eff_ham} and \eqnref{eq:overlap}. (A similar equation was derived in
the DMRG context in Ref.~\cite{dorando2009art}. See also Ref.~\cite{ueda2006lap}.)
\eqnref{eq:tdhft} and \eqnref{eq:mps_dynamics} can be formally solved with a
time-ordered exponential, which for the MPS takes the form
\begin{equation}
	\mat{A}(t) =
	\mathcal{T} \exp \left\{
		\dfrac{1}{i\hbar} \int dt \ \mat{S}^{-1}(t) \cdot \mat{H}(t)
	\right\} \cdot \mat{A}(0) .
	\label{eq:mps_exponential}
\end{equation}

Moreover, the equations of motion for the MPS can be efficiently propagated. For
example, if the time interval is discretized into units of duration $\Delta t$,
then \eqnref{eq:mps_dynamics} can be used to obtain the MPS at $t_{n+1}$ from
the MPS at $t_n$. Defining $\epsilon = \Delta t / i \hbar$,
\eqnref{eq:mps_dynamics} gives
\begin{equation}
	\mat{S}_n \cdot \Delta \vec{A}_n = \epsilon \, \vec{B}_n ,
	\label{eq:finite_difference}
\end{equation}
where $\vec{B}_n = \mat{H}_n \cdot \vec{A}_n$. This is a linear equation which
can be solved iteratively with complexity $\mathcal{O}(k M^3)$. In practice more
sophisticated time-propagation schemes, such as norm and energy conserving
propagators similar to those used in time-dependent Hartree-Fock theory, should
be employed.

\mysec{Connections to Time Evolution Algorithms}
The fundamental difficulty in simulating the evolution of a matrix product state
is that the matrix dimension required to faithfully represent the exact wave
function grows exponentially in time. The Lagrangian approach described above
leads to a set of analytic equations for the evolution of an MPS wave function
whose auxiliary dimension \emph{is fixed at $M$ throughout the simulation}. In
contrast, existing methods for the time evolution of MPS allow the dimension of
the matrix to first grow at each time step, then project back onto an MPS of
auxiliary dimension $M$. We now examine the approximate projections in existing
time evolution algorithms, and show that the analytic approach described above
is an \emph{optimal} projection in the limit of an infinitesimal time step.

In time evolution by block decimation (TEBD) \cite{vidal2004tebd} and the
time-dependent density matrix renormalization group (t-DMRG)
\cite{white2004tdmrg, daley2004tdmrg, schmitteckert2004net}, the evolution
operator $\exp (\epsilon \mathcal{H})$ is factored into a product of local
operators using a Trotter decomposition. Each local evolution operator is then
applied in sequence. A two-site evolution operator $\mathcal{U}_{i,i+1}$ between
neighboring sites joins two tensors of the MPS, $\mat{A}_i$ and $\mat{A}_{i+1}$,
into a single object, $\mat{T}_{i,i+1}$. The tensor $\mat{T}$ is then
approximately projected into a product of two tensors of the same size as
$\mat{A}_i$ and $\mat{A}_{i+1}$ through the singular value decomposition. For a
single time step (sweeping over all evolution operators for a local Hamiltonian)
this update is efficient and is $\mathcal{O}(k M^3)$. However, because the
approximate projection is only performed locally, rather than involving all the
tensors, the updated matrix product state is not the best representation of the
evolved wave function for a given auxiliary dimension. Furthermore, the
algorithm is incompatible with long range Hamiltonians.

In time-dependent matrix product states (tMPS) \cite{verstraete2004tmps}, the
full evolution operator is applied to the current state before the projection
onto an MPS of auxiliary dimension $M$. In the projection, tMPS attempts to
minimize the cost function 
\begin{equation}
	\Delta[\ket{\psi_{n+1}}]
	= \left\|
		\ket{\psi_{n+1}} - e^{\epsilon \mathcal{H}} \ket{\psi_n}
	\right\|^2 \label{eq:tmps_cost}
\end{equation}
where $\ket{\psi_{n+1}}$ and $\ket{\psi_n}$ are the new and old MPS
respectively. The minimization is of complexity $\mathcal{O}(k M^3)$, and
yields in principle the best projected MPS wave function. However, since
$\ket{\psi_{n+1}}$ depends non-linearly on $\mat{A}_i$, the minimization is not
guaranteed to find the optimal solution in practice. 

For small time step, however, the projection can be done \emph{without} any
non-linear minimization. We see this by recognizing that
\begin{equation}
i \hbar \, \epsilon \frac{\partial \ket{\psi_n}}{\partial
t}=\ket{\psi_{n+1}}-\ket{\psi_n}+O(dt^2)
\end{equation}
Substituting into the cost function \eqnref{eq:tmps_cost} and minimizing with
respect to changes in $\vec{A}$ yields the linear discretized equation of motion
obtained in \eqnref{eq:finite_difference}. Thus propagation of the MPS equation
of motion exactly determines the optimal projection in a tMPS algorithm, without
a nonlinear minimization, in the limit $\Delta t \rightarrow 0$.

\mysec{Random Phase Approximation}
Excited state properties, such as the spectrum and other expectation values, can
be obtained without studying the full time evolution of the system. Only the
linearized time evolution, or linear response, need be considered. In TDHF,
linearization of the equations of motion around a stationary state leads to the
random phase approximation. This is achieved in \eqnref{eq:tdhft} by taking
$\mat{C}(t) = \mat{C}_0 + \mat{D}(t)$ and expanding all quantities to linear
order. We now show that a similar approach to the MPS equations of motion in
\eqnref{eq:mps_dynamics} yields an MPS analog of the RPA, from which quantities
such as the excitation spectrum may be obtained. These results expand on the
analytic response theory we recently described in the DMRG context in
Ref.~\cite{dorando2009art}.

We take the zeroth-order MPS to be the ground state, defined by $\vec{A}$, with
energy $E_0$. The time-dependent MPS is defined by
\begin{equation}
	\vec{A}(t) = \vec{A} + \vec{b}(t) .
\end{equation}
The wave function and its derivatives are given by
\begin{align}
	\ket{\psi} &= \dfrac{1}{k} \sum_\mu 
	\big[ \vec{A}_\mu + \vec{b}_\mu (t) \big] \ket{\psi_{\mu}^{(0)}} \\
	\ket{\psi_\mu} &= \dfrac{1}{k} \sum_\nu 
	\big[ \vec{A}_\nu + \vec{b}_\nu (t) \big] \ket{\psi_{\mu\nu}^{(0)}} ,
\end{align}
where $\ket{\psi^{(0)}}$ and its derivatives are evaluated with $\vec{b} = 0$.

Expanding \eqnref{eq:mps_dynamics} to first order in $\vec{b}$ gives
\begin{equation}
	i \hbar \, \mat{S} \dfrac{ d \vec{b} }{ d t }
	= E_0 \, \mat{S} \cdot \vec{A}
	+ \mat{H} \cdot \vec{b} + \mat{W} \cdot \vec{b}^*.
	\label{eq:mps_rpa}
\end{equation}
The matrices $\mat{S}$ and $\mat{H}$ are the zeroth-order overlap and
Hamiltonian matrices. The matrix $\mat{W}$ couples $\vec{b}$ to its complex
conjugate and is defined by
 \begin{equation}
 	\mat{W}_{\mu\nu} = \dfrac{1}{k}
 	\bra{\psi_{\mu\nu}^{(0)}} \mathcal{H} \ket{\psi^{(0)}}.
 \end{equation}
$\mat{W}$ is symmetric, but not Hermitian.

The first term on the right-hand-side of \eqnref{eq:mps_rpa} can be eliminated
by multiplying the entire wave function $\ket{\psi}$ by the phase factor $e^{-
i E_0 t / \hbar}$. The equations of motion for $\vec{b}(t)$ are harmonic and may
be solved by taking $ \vec{b}(t) = \vec{X} \, e^{i \omega t} + \vec{Y}^* \,
e^{-i \omega t}$. \eqnref{eq:mps_rpa} then gives
\begin{equation}
	\hbar \omega \, \begin{bmatrix} \mat{S} & 0 \\ 0 & -\mat{S}^* \end{bmatrix}
	\begin{pmatrix} \vec{X} \\ \vec{Y} \end{pmatrix}
	= \begin{bmatrix} \mat{H} & \mat{W} \\ \mat{W}^* & \mat{H}^* \end{bmatrix}
	\begin{pmatrix} \vec{X} \\ \vec{Y} \end{pmatrix} .
\end{equation}
The eigenvectors of this system of equations define the normal modes of the
system, and the (positive) eigenvalues approximate its excitation spectrum. They
are both efficiently obtained with $\mathcal{O}(kM^3)$ complexity. The normal
modes define the response matrix $\Pi(\omega)$, which determines all the
response properties of the system:
\begin{equation}
	\Pi(\omega) = \sum_q
	\begin{pmatrix} \vec{X}_{q}^{\,} \\ \vec{Y}_{q}^{\,} \end{pmatrix}
	\dfrac{1}{\omega - \omega_q}
	\begin{pmatrix} \vec{X}_{q}^{*} & \vec{Y}_{q}^{*} \end{pmatrix} .
	\label{eq:response_matrix}
\end{equation}
For example, consider a harmonic perturbation, which defines a source
$\vec{q}(t)$ for $\vec{b}$:
\begin{equation}
	\vec{q}{}_\mu = \bra{ \psi^{(0)}_{\mu} } \mathcal{Q}(t) \ket{ \psi^{(0)} } .
\end{equation}
(The expectation of $\mathcal{Q}(t)$ in the ground state is assumed to vanish.)
Using $\Pi(\omega)$, the time-dependent variation in an observable $\mathcal{P}$
due to $\mathcal{Q}(t)$ is given by
\begin{equation}
	\langle \delta \mathcal{P}(\omega) \rangle
	= \begin{bmatrix} \vec{p}^*(\omega) & \vec{p}(\omega) \end{bmatrix}
	\cdot \Pi(\omega) \cdot
	\begin{bmatrix} \vec{q}(\omega) \\ \vec{q}^{*}(\omega) \end{bmatrix}
\end{equation}

The linearized equations of motion describe a superposition of MPS wave
functions. Each tensor product $\mat{A}_1 \cdot \mat{A}_2 \dots \mat{A}_N $ is
replaced by a sum of tensor products:
\begin{align*}
	\mat{A}_1 \cdot \mat{A}_2 \dots \mat{A}_N \
	&+ \ \delta\mat{A}_1 \cdot \mat{A}_2 \dots \mat{A}_N \\
	+ \ \mat{A}_1 \cdot \delta\mat{A}_2 \dots \mat{A}_N \
	+ &\ \cdots \ +
	\ \mat{A}_1 \cdot \mat{A}_2 \dots \delta\mat{A}_N \label{eq:rpa_ansatz}
\end{align*}
Thus, the analog of the RPA uses an MPS with auxiliary dimension $M$ to give the
response properties of an MPS of much larger dimension $kM$. 

\mysec{Green functions}
HF theory is an independent particle mean-field theory and the natural Green
functions that emerge are labeled by particle indices. MPS are based on a
site-based mean field theory and the natural Green functions that emerge are
labeled by sites. Here we define site-based Green functions, show how they are
obtained in the RPA formalism above, and analyze the nature of the correlations
they contain.

First, we define the one-site density matrix $\Gamma^{(i)}$, which is the
site-based analog of the one-particle density matrix. This is obtained from
$|\psi\rangle\langle \psi|$ by tracing out all sites of the lattice except site
$i$. The expectation value of a one-site operator $\mathcal{P}^{(i)}$ is given
by the trace
\begin{equation}
	\langle \mathcal{P}^{(i)} \rangle = \sum_{nn'} \Gamma^{(i)}_{nn'} \mat{P}^{(i)}_{nn'} .
\end{equation}
Each element $\Gamma^{(i)}_{nn'}$ is an expectation value of an operator
$\hat{\gamma}_{nn'}$ defined by
\begin{equation}
	\bra{\bar{n}} \hat{\gamma}_{nn'} \ket{\bar{n}'} = \delta_{n\bar{n}} \delta_{n'\bar{n}'} .
\end{equation}
For a system with 2 physical degrees of freedom per site, the operators
$a_{i}^{\,}$ and $ a_{i}^{\dag}$ correspond to $\hat{\gamma}_{01}$ and
$\hat{\gamma}_{10}$, and the number operator is $\hat{\gamma}_{11}$. Creation,
annihilation, and number operators of systems with more degrees of freedom per
site can be constructed from $\hat{\gamma}$ as well. Thus the one-site density
matrix contains components of one-, two-, and mixed-particle density matrices.

We now define a site-site (retarded) Green function $\Pi^{(ij)}(t)$. We
consider the response of the site density matrix at site $i$ and time $t$,
$\Gamma^{(i)}(t)$, to a perturbation at site $j$,
$\mathcal{Q}^{(j)} = \delta(t) \, \sum_{nn^\prime} V_{nn^\prime}
\hat{\gamma}^{(j)}_{nn^\prime}$. We then have
\begin{equation}
	\Pi_{nn';\bar{n}\bar{n}'}^{(ij)}(t) =
	\dfrac{\partial \Gamma^{(i)}_{nn'}(t)}{\partial V_{\bar{n}\bar{n}'}}.
\end{equation}
Note that by choosing appropriate combinations of $\gamma^{(i)}_{nn'}$ and
$\gamma^{(j)}_{\bar{n} \bar{n}'}$, it is possible to construct spectral
functions and one-, two-particle, and other types of conventional Green
functions.

Within the RPA framework, the elements of $\Pi^{(ij)}$ are obtained from the MPS
response matrix $\Pi_{\mu\nu}$ as
\begin{equation}
	\Pi_{nn';\bar{n}\bar{n}'}^{(ij)}(\omega) =
	\bra{\psi_\mu} \gamma^{(i)}_{nn'} \ket{\psi} \, 
	\Pi_{\mu\nu}(\omega) \,
	\bra{\psi} \gamma^{(j)}_{\bar{n}\bar{n}'} \ket{\psi_\nu}.
\end{equation}

The response function and the correlation functions of the ground state are
related through the fluctuation-dissipation theorem \cite{callen1951fdt,
pines1994toq}. This allows us to analyze the nature of the additional
correlations introduced into the MPS at the RPA level. We have
\begin{equation}
	\langle \gamma_{nn'}^{(i)} \, \gamma_{\bar{n}\bar{n}'}^{(j)} \rangle
	= - \dfrac{1}{2\pi} \int_{-\infty}^{+\infty} d\omega \,
	\Pi_{nn';\bar{n}\bar{n}'}^{(ij)}(\omega)
\end{equation}
From the expression for the response matrix $\Pi_{\mu\nu}(\omega)$ above, we see
that the correlation function is composed of a sum of terms, each of the form
\begin{align}
 	\sum_{q,\mu\nu} \bra{\psi} \gamma^{(j)}_{\bar{n}\bar{n}'} \ket{\psi_\nu}
	X_{q\nu}^{*} & X_{q\mu}^{\,} \bra{\psi_\mu} \gamma^{(i)}_{nn'} \ket{\psi}
	\notag \\
 	&= \bra{\psi} \gamma^{(j)}_{\bar{n}\bar{n}'} \cdot \Lambda_{XX} \cdot
 	\gamma^{(i)}_{nn'} \ket{\psi}
\end{align}
plus the corresponding contributions from $\Lambda_{XY}$, $\Lambda_{YX}$, and
$\Lambda_{YY}$ from \eqnref{eq:response_matrix}. $\Lambda_{XX}$ is a matrix
product operator (MPO), and is responsible for additional correlation introduced
at the RPA level relative to the ground state (which is obtained by setting all
$\Lambda = 1$). By regrouping terms in the sum, we obtain
\begin{equation}
	\Lambda_{XX} = \sum_{q}
	\Big( \sum_i \mat{X}_{qi}^{I*} \ket{\psi_I} \Big)
	\Big(\sum_j \bra{\psi_J} \mat{X}_{qj}^{J} \Big)
\end{equation}
Each term in parentheses is a sum of $k$ MPS of dimension $M$, and their outer
product is formally an MPO of dimension $2kM$. The sum includes an MPO for each
normal mode, making $\Lambda_{XX}$ an MPO of very large auxiliary dimension.
Thus, the entanglement and correlation possible at the RPA level are greater
than the auxiliary dimension $M$ of the matrices might suggest.

\mysec{Conclusions}
In this paper, we have drawn on parallels between the product structure of HF
theory and MPS to develop analytic equations of motion for time evolution, an
MPS RPA for response and excitations, and a theory of site-based Greens
functions which introduce correlations beyond the MPS ground state, as
demonstrated via the fluctuation-dissipation theorem. The new time evolution
and response algorithms proposed here can all be implemented with a complexity
proportional to the ground state DMRG algorithm. Finally, our work suggests
that further extensions of Hartree-Fock theory may also be carried over to MPS,
and these will be explored in the future.

This work was supported by the
	the Cornell Center for Materials Research,
	the Center for Molecular Interfacing,
	NSF CAREER,
	DOE CSGF,
	the Camille and Henry Dreyfus Foundation,
	the David and Lucile Packard Foundation,
	and the Alfred P. Sloan Foundation.


\begin{thebibliography}{17}
\expandafter\ifx\csname natexlab\endcsname\relax\def\natexlab#1{#1}\fi
\expandafter\ifx\csname bibnamefont\endcsname\relax
  \def\bibnamefont#1{#1}\fi
\expandafter\ifx\csname bibfnamefont\endcsname\relax
  \def\bibfnamefont#1{#1}\fi
\expandafter\ifx\csname citenamefont\endcsname\relax
  \def\citenamefont#1{#1}\fi
\expandafter\ifx\csname url\endcsname\relax
  \def\url#1{\texttt{#1}}\fi
\expandafter\ifx\csname urlprefix\endcsname\relax\def\urlprefix{URL }\fi
\providecommand{\bibinfo}[2]{#2}
\providecommand{\eprint}[2][]{\url{#2}}

\bibitem[{\citenamefont{White}(1993)}]{white1993dmrg}
\bibinfo{author}{\bibfnamefont{S.}~\bibnamefont{White}},
  \bibinfo{journal}{Phys. Rev. B} \textbf{\bibinfo{volume}{48}},
  \bibinfo{pages}{10345} (\bibinfo{year}{1993}).

\bibitem[{\citenamefont{Rommer and {\"O}stlund}(1997)}]{rommer1997mps}
\bibinfo{author}{\bibfnamefont{S.}~\bibnamefont{Rommer}} \bibnamefont{and}
  \bibinfo{author}{\bibfnamefont{S.}~\bibnamefont{{\"O}stlund}},
  \bibinfo{journal}{Phys. Rev. B} \textbf{\bibinfo{volume}{55}},
  \bibinfo{pages}{2164} (\bibinfo{year}{1997}).

\bibitem[{\citenamefont{McLachlan and Ball}(1964)}]{mclachlan1964tdhft}
\bibinfo{author}{\bibfnamefont{A.}~\bibnamefont{McLachlan}} \bibnamefont{and}
  \bibinfo{author}{\bibfnamefont{M.}~\bibnamefont{Ball}},
  \bibinfo{journal}{Rev. Mod. Phys.} \textbf{\bibinfo{volume}{36}},
  \bibinfo{pages}{844} (\bibinfo{year}{1964}).

\bibitem[{\citenamefont{Luo et~al.}(2003)\citenamefont{Luo, Xiang, and
  Wang}}]{xiang2003tddmrg}
\bibinfo{author}{\bibfnamefont{H.~G.} \bibnamefont{Luo}},
  \bibinfo{author}{\bibfnamefont{T.}~\bibnamefont{Xiang}}, \bibnamefont{and}
  \bibinfo{author}{\bibfnamefont{X.~Q.} \bibnamefont{Wang}},
  \bibinfo{journal}{Phys. Rev. Lett.} \textbf{\bibinfo{volume}{91}},
  \bibinfo{pages}{049701} (\bibinfo{year}{2003}).

\bibitem[{\citenamefont{Vidal}(2004)}]{vidal2004tebd}
\bibinfo{author}{\bibfnamefont{G.}~\bibnamefont{Vidal}},
  \bibinfo{journal}{Phys. Rev. Lett.} \textbf{\bibinfo{volume}{93}},
  \bibinfo{pages}{40502} (\bibinfo{year}{2004}).

\bibitem[{\citenamefont{White and Feiguin}(2004)}]{white2004tdmrg}
\bibinfo{author}{\bibfnamefont{S.}~\bibnamefont{White}} \bibnamefont{and}
  \bibinfo{author}{\bibfnamefont{A.}~\bibnamefont{Feiguin}},
  \bibinfo{journal}{Phys. Rev. Lett.} \textbf{\bibinfo{volume}{93}},
  \bibinfo{pages}{76401} (\bibinfo{year}{2004}).

\bibitem[{\citenamefont{Daley et~al.}(2004)\citenamefont{Daley, Kollath,
  Schollw{\"o}ck, and Vidal}}]{daley2004tdmrg}
\bibinfo{author}{\bibfnamefont{A.}~\bibnamefont{Daley}},
  \bibinfo{author}{\bibfnamefont{C.}~\bibnamefont{Kollath}},
  \bibinfo{author}{\bibfnamefont{U.}~\bibnamefont{Schollw{\"o}ck}},
  \bibnamefont{and} \bibinfo{author}{\bibfnamefont{G.}~\bibnamefont{Vidal}},
  \bibinfo{journal}{J. Stat. Mech.} \textbf{\bibinfo{volume}{2004}},
  \bibinfo{pages}{P04005} (\bibinfo{year}{2004}).

\bibitem[{\citenamefont{Schmitteckert}(2004)}]{schmitteckert2004net}
\bibinfo{author}{\bibfnamefont{P.}~\bibnamefont{Schmitteckert}},
  \bibinfo{journal}{Phys. Rev. B} \textbf{\bibinfo{volume}{70}},
  \bibinfo{pages}{121302(R)} (\bibinfo{year}{2004}).

\bibitem[{\citenamefont{Verstraete et~al.}(2004)\citenamefont{Verstraete,
  Garcia-Ripoll, and Cirac}}]{verstraete2004tmps}
\bibinfo{author}{\bibfnamefont{F.}~\bibnamefont{Verstraete}},
  \bibinfo{author}{\bibfnamefont{J.}~\bibnamefont{Garcia-Ripoll}},
  \bibnamefont{and} \bibinfo{author}{\bibfnamefont{J.}~\bibnamefont{Cirac}},
  \bibinfo{journal}{Phys. Rev. Lett.} \textbf{\bibinfo{volume}{93}},
  \bibinfo{pages}{207204} (\bibinfo{year}{2004}).

\bibitem[{\citenamefont{Schollw{\"o}ck}(2010)}]{schollwock2010dmrg}
\bibinfo{author}{\bibfnamefont{U.}~\bibnamefont{Schollw{\"o}ck}},
  \bibinfo{journal}{Ann. Phys.} \textbf{\bibinfo{volume}{326}},
  \bibinfo{pages}{96} (\bibinfo{year}{2010}).

\bibitem[{\citenamefont{Hallberg}(1995)}]{hallberg1995dma}
\bibinfo{author}{\bibfnamefont{K.}~\bibnamefont{Hallberg}},
  \bibinfo{journal}{Phys. Rev. B} \textbf{\bibinfo{volume}{52}},
  \bibinfo{pages}{9827} (\bibinfo{year}{1995}).

\bibitem[{\citenamefont{K{\"u}hner and White}(1999)}]{kuhner1999dcf}
\bibinfo{author}{\bibfnamefont{T.}~\bibnamefont{K{\"u}hner}} \bibnamefont{and}
  \bibinfo{author}{\bibfnamefont{S.}~\bibnamefont{White}},
  \bibinfo{journal}{Phys. Rev. B} \textbf{\bibinfo{volume}{60}},
  \bibinfo{pages}{335} (\bibinfo{year}{1999}).

\bibitem[{\citenamefont{Jeckelmann}(2002)}]{jeckelmann2002ddmrg}
\bibinfo{author}{\bibfnamefont{E.}~\bibnamefont{Jeckelmann}},
  \bibinfo{journal}{Phys. Rev. B} \textbf{\bibinfo{volume}{66}},
  \bibinfo{pages}{45114} (\bibinfo{year}{2002}).

\bibitem[{\citenamefont{Dorando et~al.}(2009)\citenamefont{Dorando, Hachmann,
  and Chan}}]{dorando2009art}
\bibinfo{author}{\bibfnamefont{J.}~\bibnamefont{Dorando}},
  \bibinfo{author}{\bibfnamefont{J.}~\bibnamefont{Hachmann}}, \bibnamefont{and}
  \bibinfo{author}{\bibfnamefont{G.}~\bibnamefont{Chan}}, \bibinfo{journal}{J.
  Chem. Phys.} \textbf{\bibinfo{volume}{130}}, \bibinfo{pages}{184111}
  (\bibinfo{year}{2009}).

\bibitem[{\citenamefont{{Ueda} et~al.}(2006)\citenamefont{{Ueda}, {Jin},
  {Shibata}, {Hieida}, and {Nishino}}}]{ueda2006lap}
\bibinfo{author}{\bibfnamefont{K.}~\bibnamefont{{Ueda}}},
  \bibinfo{author}{\bibfnamefont{C.}~\bibnamefont{{Jin}}},
  \bibinfo{author}{\bibfnamefont{N.}~\bibnamefont{{Shibata}}},
  \bibinfo{author}{\bibfnamefont{Y.}~\bibnamefont{{Hieida}}}, \bibnamefont{and}
  \bibinfo{author}{\bibfnamefont{T.}~\bibnamefont{{Nishino}}},
  \eprint{arXiv:cond-mat/0612480},
  (\bibinfo{year}{2006}).

\bibitem[{\citenamefont{Callen and Welton}(1951)}]{callen1951fdt}
\bibinfo{author}{\bibfnamefont{H.}~\bibnamefont{Callen}} \bibnamefont{and}
  \bibinfo{author}{\bibfnamefont{T.}~\bibnamefont{Welton}},
  \bibinfo{journal}{Phys. Rev.} \textbf{\bibinfo{volume}{83}},
  \bibinfo{pages}{34} (\bibinfo{year}{1951}).

\bibitem[{\citenamefont{Pines and Nozieres}(1994)}]{pines1994toq}
\bibinfo{author}{\bibfnamefont{D.}~\bibnamefont{Pines}} \bibnamefont{and}
  \bibinfo{author}{\bibfnamefont{P.}~\bibnamefont{Nozieres}},
  \emph{\bibinfo{title}{{Theory of Quantum Liquids: Normal Fermi Liquids}}}
  (\bibinfo{publisher}{Westview Press}, \bibinfo{year}{1994}).

\end{thebibliography}
\end{document}